\documentclass[final,3p,times, twocolumn]{elsarticle}

\usepackage{graphicx}

\usepackage{amssymb,amsmath}
\usepackage{amsthm}
\usepackage{natbib}
\usepackage{algorithm}
\usepackage[noend]{algpseudocode}
\usepackage{algorithmicx}
\usepackage{url}

\usepackage{subfig}

\begin{document}

\begin{frontmatter}







\title{Joint Spatial Multiplexing and Transmit Diversity in MIMO Ad Hoc Networks}

\author[tcd]{Fadhil Firyaguna}

\author[unb]{Ana C. O. Christ\'ofaro}
\author[unb]{\'Everton A. L. Andrade}
\author[unb]{Tiago S. Bonfim}
\author[unb]{Marcelo M. Carvalho\corref{cor1}}
\ead{mmcarvalho@ene.unb.br}

\address[unb]{Department of Electrical Engineering, University of Bras\' ilia,  Brazil}
\address[tcd]{CONNECT Centre, Trinity College Dublin, Ireland}

\cortext[cor1]{Corresponding author}

\begin{abstract}
This paper investigates the performance of MIMO ad hoc networks that employ transmit diversity, as delivered by the Alamouti scheme, and/or spatial multiplexing, according to the Vertical Bell Labs Layered Space-Time system (V-BLAST). Both techniques are implemented in a discrete-event network simulator by focusing on their overall effect on the resulting signal-to-interference-plus-noise ratio (SINR) at the intended receiver. 
Unlike previous works that have studied fully-connected scenarios or have assumed simple abstractions to represent MIMO behavior, this paper evaluates MIMO ad hoc networks that are {\it not} fully connected by taking into account the effects of multiple antennas on the clear channel assessment (CCA) mechanism of CSMA-like medium access control (MAC) protocols. In addition to presenting a performance evaluation of ad hoc networks operating according to each individual MIMO scheme, this paper proposes simple modifications to the IEEE 802.11 DCF MAC to allow the joint operation of both MIMO techniques. Hence, each pair of nodes is allowed to select the best MIMO configuration for the impending data transfer. The joint operation is based on three operation modes that are selected based on the estimated SINR at the intended receiver and its comparision with a set of threshold values. The performance of ad hoc networks operating with the joint MIMO scheme is compared with their operation using each individual MIMO scheme and the standard SISO IEEE 802.11. Performance results are presented based on MAC-level throughput per node, delay, and fairness under saturated traffic conditions.

\end{abstract}

\begin{keyword}
MIMO networks \sep  ad hoc networks \sep medium access control protocols \sep wireless networks 
\end{keyword}

\end{frontmatter}


\section{Introduction}

\label{sec:introduction}


In the past decade, the world has witnessed the widespread adoption of {\it multiple-input multiple-output} (MIMO) technologies in a number of wireless systems and devices. To date, many standards on wireless communications have already embraced MIMO as a key technology to allow high spatial reuse, transmission rate, and/or strong resilience to channel impairments~\cite{paul.circuitsandsystems.magazine.2008}. However, while the use of MIMO systems in infrastructure-based wireless networks is already a reality, its adoption in infrastructure-less networks is not as well mature and understood. In fact, as far as research on multi-antenna ad hoc networks is concerned, a significant body of work has already been done on the use of beamforming techniques (i.e., the so-called directional antennas)~\cite{choudhury.icnp04, ramanathan.jsac05}. However, the study of MIMO ad hoc networks that exploit {\it diversity} and/or {\it multiplexing} gains still deserves further investigation, since the bulk of previous work has either focused on fully-connected and collision-free scenarios~\cite{hu.jcn04,ieee.levorato.casari.2007, rosseto.trans.wireless.comm.2009,babich.iet.2014}, or have assumed simple abstractions to represent MIMO behavior at the PHY layer~\cite{sundaresan.icnp.2005, jakllari.infocom.2006, gelal.diversity.06}. 

In practice, the multiple access interference (MAI) that results from the dynamics of the underlying medium access control (MAC) protocol under a radio-based topology may severely compromise the expected gains of MIMO technologies in ad hoc networks. Surprisingly, very few works have addressed the impact of MAI on the performance of ad hoc networks when nodes are not fully connected and employ diversity and/or spatial multiplexing. Under such scenarios, and assuming a CSMA-like MAC protocol, hidden-terminals become apparent, and issues related to clear channel assessment (CCA) become key to overall network performance, since spatial reuse may either be inhibited or enhanced, depending on how physical and virtual carrier sensing are carried out with the set of available antennas. As a result, the assessment of network-wide impact of MIMO solutions in ad hoc networks still requires further investigation, since the aforementioned issues may compromise overall performance. 
In particular, there is still a need for further studies on network performance when diversity and/or spatial multiplexing are deployed in MIMO ad hoc networks not fully connected.  

To shed some light on this problem, this paper investigates the performance of  MIMO ad hoc networks that are {\it not} fully connected, and whose nodes utilize {\it transmit diversity}, as delivered by the Alamouti scheme~\cite{alamouti}, and/or {\it spatial multiplexing}, according to the Vertical Bell Labs Layered Space-Time system (V-BLAST)~\cite{ieee.foshini.ursi.1998}. Both techniques are implemented in the ns-3 discrete-event network simulator~\cite{ns-3} by focusing on the overall effects that each of these MIMO technologies cause on the resulting signal-to-interference-plus-noise (SINR) at the intended receiver. 
Consequently, as opposed to previous works~\cite{sundaresan.icnp.2005, jakllari.infocom.2006, gelal.diversity.06}, this work does not simply assume constant gains to abstract MIMO performance at the PHY layer. Instead, the effects of MAI are taken into account according to the underlying MIMO channel model and network topology. Moreover, because the focus of this paper is on CSMA-based MIMO ad hoc networks, the issue of clear channel assessment (CCA) under multiple antennas is also investigated. Both physical and virtual carrier sensing are treated by considering the IEEE 802.11 DCF MAC as the MAC of choice. 

Based on a preliminary evaluation of the performance of each MIMO scheme under different antenna configurations (both Alamouti and V-BLAST), this paper also proposes simple modifications to the IEEE 802.11 DCF MAC to allow the joint operation of both MIMO techniques, i.e., to allow 
each pair of nodes to select the best MIMO configuration for a particular data transfer. Following the findings of our performance study, the proposed joint operation is based on three transmission modes: {\it full transmit diversity}, {\it spatial multiplexing with diversity}, and {\it full spatial multiplexing}. The selection of a given transmit mode is based on a simple switching mechanism that takes into account the SINR estimation at the intended receiver and corresponding comparison with a set of threshold values found with the help of another performance study. Finally, the performance of the joint MIMO scheme is compared with each individual MIMO configuration and with the standard SISO IEEE 802.11 DCF MAC in terms of MAC-level throughtput per node,  delay, and fairness under saturated traffic conditions.

The rest of the paper is organized as follows. Section~\ref{sec:related work} discusses related work. Section~\ref{sec:mimo_models} presents the MIMO channel model along with a short description of the Alamouti and V-BLAST MIMO systems. Section~\ref{sec:mimo_at_mac} describes the needed modifications to the MAC layer for proper operation of the MIMO ad hoc network. Section~\ref{sec:individual_mimo_schemes} presents the performance of ad hoc networks enabled with each individual MIMO scheme, and Section~\ref{sec:joint-mimo} contains the proposal for joint operation of both schemes. Finally, Section~\ref{sec:conclusions} contains the conclusions.


\section{Related Work}\label{sec:related work}


As far as the application of MIMO systems to exploit diversity and/or multiplexing gains, Stamoulis and Al-Dhahir~\cite{stamoulis.transwirel03} have investigated the impact of space-time block codes (STBC) on IEEE 802.11a WLANs operating in ad hoc mode. They have used 
packet traces in the ns-2 simulator to evaluate the benefits of STBC on the performance of upper-layer protocols, such as TCP. They have assumed fully-connected networks with the simplest $2 \times 1$ Alamouti scheme. Later, Hu and Zhang~\cite{hu.jcn04} have attempted to model MIMO ad hoc networks by focusing on IEEE 802.11 with STBC. Their modeling approach disregards the impact of network topology by assuming that events experienced by one station are statistically the same as those of other stations. Therefore, in practice, each node is treated as surrounded by the same average number of nodes, and a multihop network is simplified to many single-hop networks, where interactions occur only with immediate neighbors. 

Rosseto and Zorzi~\cite{rosseto.trans.wireless.comm.2009} have studied the use of a class of STBC to transmit control frames as a means to provide range extension and omnidirectionality. In their study, they have only considered single-hop networks. Consequently, they have not evaluated the impact of range extension (via STBC) on the carrier sense activity of networks not fully connected (under hidden terminals). In addition, their channel model neglects the interference from other users (i.e., no MAI), which leads to overestimation of network throughput. Likewise, Levorato {\it et. al.}~\cite{ieee.levorato.casari.2007} have also considered fully-connected networks to evaluate the performance of MIMO-BLAST ad hoc networks. Their work presents a physical layer abstraction based on an analytical model that predicts the error propagation of symbol detection during the V-BLAST iterative mechanism. Along the same lines, Gelal~\textit{et al.}~\cite{gelal.diversity.06} have examined the diversity gains delivered by STBC to exploit range extension and rate increase. Their PHY-layer model assumes high SNR regime and a {\it constant} diversity gain of about 15 dB to represent the diversity gain of a  $4 \times 1$ STBC MIMO. Such an assumption oversimplifies the analysis and, unfortunately, precludes the scenarios where STBC is most needed (and it shows its actual strength): when links present low SINR. 

Xi {\it et al.}~\cite{xi.opnet} have proposed an open-loop link adaptation algorithm for IEEE 802.11n MIMO networks based on SNR information estimated from a channel model and the most recent received packet. Their algorithm adapts both modulation and coding scheme (MCS), as well as the MIMO mode. 
But, in order to focus on the impact of transmission modes on the achievable throughput, their work disregards the impact of {\it frame collisions} on throughput evaluations. In simulations, their algorithm is evaluated in network topologies with only six nodes. 
Likewise, Xia {\it et al.}~\cite{arfht.xia.hamdi} have proposed an open-loop autorate fallback algorithm for IEEE 802.11n MIMO networks. The algorithm estimates link quality in order to select the most appropriate transmission scheme (both MCS and MIMO mode). The feedback takes into account ACK frames and the received signal strength measured at each receive antenna. 
However, their work focus on link-level performance only, without considering the impact of contention, network topology, and multiple access interference.

Siam and Krunz~\cite{siam.wireless.net.applications.2008} have proposed the combined MAC (CMAC), a power-controlled MAC protocol for MIMO ad hoc networks that allows the nodes to dinamically switch between diversity and multiplexing modes according to an utility function that relates throughput with energy consumption. CMAC uses a modified CSMA/CA where the contention period, called {\it access window} (AW), consists of a variable number of fixed-duration {\it access slots} (AS) used by a pair of nodes to agree on communication parameters. 
All data transfers scheduled in each AS happen concurrently, in the end of the AW. In CMAC, the first RTS in an AW is sent at {\it maximum power} in order to ensure the farthest transmission range for this RTS. Although it is argued that the negative impact of sending an RTS at maximum power is compensated by concurrent transmissions, CMAC is only evaluated under single-hop scenarios. Therefore, its impact on spatial reuse under hidden terminals is unknown when nodes are  not within range of each other (but can be {\it sensed} from further distant nodes due to carrier sensing). 
In addition, the power computations for correct operation of CMAC disregard the impact of aggregate interference from distant nodes, and assume explicit knowledge of the exact distance between nodes and the empirical path-loss exponent of channel propagation---a parameter not usually available and estimated in real time in practical applications. Finally, CMAC delay and fairness performance are not evaluated, and MAC queues are assumed to be infinite. The same authors also propose MIMO-POWMAC~\cite{siam.infocom.2006}, a power-controlled MAC protocol for spatial multiplexing only. Similar to CMAC, the performance evaluation of of MIMO-POWMAC is carried out using single-hop networks only. To simplify their evaluation, channel gains are assumed to be symmetrical and stationary during transmission of control and data frames. 
Recently, Babich et. al.~\cite{babich.iet.2014} have presented a study on the multiplexing and diversity tradeoff in IEEE 802.11 networks through an analytical model. In their analysis, the network topology follows a destination centric approach, where both the source node and interferers are randomly located within a circle that has the destination node at its center. Consequently, their work does not address the impact of the carrier sensing range in topologies not fully-connected, where concurrent transmissions of hidden terminals and multiple access interference are a major source of errors on different source-destination pairs. In fact, they only consider the basic access mechanism and, therefore, the impact of the four-way handshake mechanism under MIMO is not addressed. In the following, we extend our preliminary works~\cite{firyaguna.iccc.2012,andrade.camad.2013} to present a more comprehensive study of individual MIMO schemes applied to ad hoc networks, as well as a proposed joint scheme based on three operation modes that are evaluated for different performance metrics in topologies not fully-connected.

\section{MIMO Models and Simulation}
\label{sec:mimo_models}

In this section we present the MIMO channel model, the MIMO transmission schemes studied in this work, and the adopted approaches to implement these models in the ns-3 discrete-event simulator~\cite{ns-3}.  

\subsection{MIMO Channel Model} 
\label{sec:mimo_channel}

We consider a MIMO ad hoc network where every node utilizes the same number $M$ of transmit antennas, and the same number $N$ of receive antennas.  It is assumed a narrowband frequency nonselective Rayleigh fading channel model, with an $N \times M$ complex baseband channel matrix $\boldsymbol{H}$ whose entries $h_{ij}$ indicate the channel gain between the $j$th transmit antenna and the $i$th receive antenna. Hence, $h_{ij} \sim \text{i.i.d. } \mathcal{CN}(0, \Omega^2)$, which means that $|h_{ij}|^2$ is exponentially distributed with average value $\Omega^2$ corresponding to the large-scale path loss {\it gain} between the sender and the receiver. In this work, we consider the two-ray path loss model~\cite{goldsmith}, and assume a total transmit power $P_t$ that is equally distributed among the $M$ antennas over a symbol period, i.e., the average power per antenna per symbol period is $P_t/M$. Each node transmits an $M \times 1$ data symbol vector $\boldsymbol{x}$ whose covariance matrix $\boldsymbol{R}_{xx} = E[\boldsymbol{x x}^H]$ is assumed to satisfy $\text{Tr}(\boldsymbol{R}_{xx}) = M$, so that the total transmit power is constrained to $P_t$.  

In an ad hoc network, any packet reception can be corrupted by noise and by potential interference caused by simultaneous transmissions from other active nodes. In particular, if a node $i$ transmits to a node $j$, and there are $K$ simultaneous transmissions over the channel, the $N \times 1$ complex baseband signal vector $\boldsymbol{y}_j$ received at node $j$, in a given time slot, will be given by 
\begin{align}
\boldsymbol{y}_j = \sqrt{\frac{P_t}{M}}\boldsymbol{H}_i^j\boldsymbol{x}_i + \sqrt{\frac{P_t}{M}}\sum_{k = 1}^{K} \boldsymbol{H}_k^j \boldsymbol{x}_k + \boldsymbol{n},
\end{align}
where $\boldsymbol{H}_k^j$ indicates the channel matrix between the source node $k$ and the destination node $j$, $\boldsymbol{x}_k$ is the symbol vector transmitted by node $k$, and $\boldsymbol{n}$ is the $N \times 1$ additive white Gaussian noise (AWGN) vector which is assumed to be $\mathcal{CN}(\boldsymbol{0}, N_0\boldsymbol{I}),$ i.e., i.i.d. in each branch. 

\subsection{The Alamouti Scheme}
\label{sec:alamouti_scheme}

The Alamouti scheme~\cite{alamouti} is a {\it transmit diversity} technique that supports maximum-likelihood detection without requiring channel state information (CSI) at the transmitter. Its operation relies on a clever transmission technique that spans over two symbol periods. Then, based only on CSI estimation at the receiver, it employs a linear transformation on the signals received over two consecutive symbol periods to effectively decouple symbol transmissions~\cite{alamouti,goldsmith}. In order to implement the Alamouti scheme in a discrete-event network simulator, we focus on its effects on the resulting SINR at a given receiver after symbol decoupling operation. Hence, assuming perfect CSI at the receiver, and the use of 2 transmit antennas and $N$ receive antennas, the resulting SINR at node $j$, for a symbol transmitted by a node $i$, is given by~\cite{mmcarvalho}  
\begin{equation}
\label{eq:sinr_alamouti} \mbox{SINR} = \frac{\|\boldsymbol{H}_i^j\|_F^2
E_s/2}{N_0 + \sum_{k=1}^{K} \|\boldsymbol{H}_k^j\|_F^2E_s/2},
\end{equation}
where $E_s$ is the transmit energy, $N_0$ is the average noise power (AWGN), $K$ is the number of active nodes leading to multiple access interference (MAI), and $\|\boldsymbol{H}_k^j\|_F^2$ is the {\em Frobenius norm} of the channel matrix $\boldsymbol{H}_k^j$ between nodes $k$ and $j$, i.e., $\|\boldsymbol{H}_k^j\|_F^2 = \sum_{n=1}^N\sum_{m=1}^2 |h_{mn}^{kj}|^2.$ 
Once the effective SINR is known, one can compute the corresponding uncoded BER for a given modulation based on mathematical formulae or lookup table.  


\subsection{The V-BLAST Scheme}
\label{sec:implementacaovblast}

\begin{figure}[htb]
\centering
\includegraphics[scale=0.35]{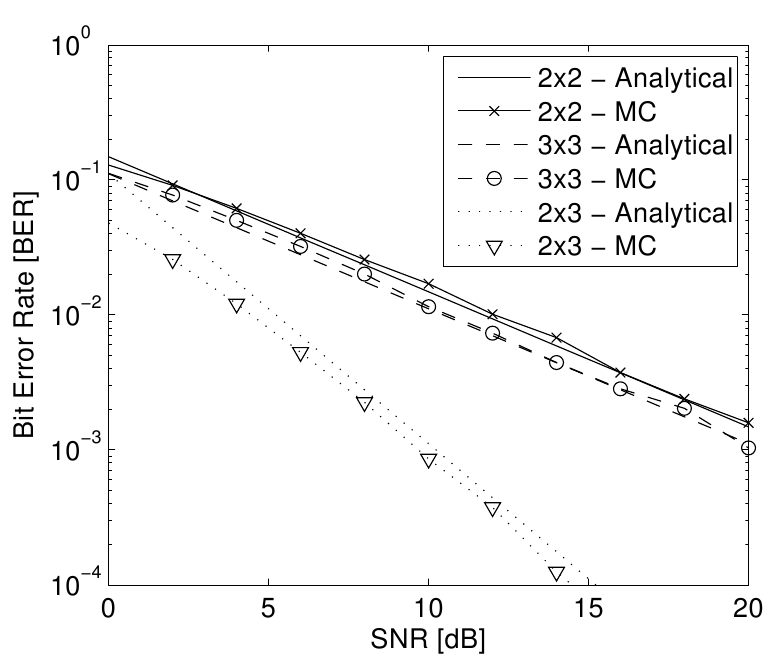}
\caption{Comparison of V-BLAST bit error rates between Loyka and Gagnon's model (``Analytical'') and Monte Carlo (``MC'') simulations.}
\label{fig:montecarlo}
\end{figure}

The Vertical Bell Labs Layered Space-Time (V-BLAST) system~\cite{ieee.foshini.ursi.1998} is a {\it spatial multiplexing} technique that splits the information bit stream into $M$ substreams for parallel transmission using the $M$ transmit antennas at the same time and frequency. At the receiver, each of the $N$ antennas capture the mixed transmitted signals, and V-BLAST processes the received signals in such a way that the channel matrix is transformed into a set of virtual parallel independent channels. Loyka and Gagnon~\cite{2006.loyka.vblast.ieee.trans.comm} have provided a performance analysis of V-BLAST based on the Gram-Schmidt orthogonalization process and a statistical analysis of the postprocessing signal-to-noise ratio for an $M \times N$ system. They have derived closed-form expressions for the average BER of uncoded $M \times N$ BPSK system. According to their results, the average total BER evaluated across all antennas (and represented in their work by $\overline{P_{et}}$) can be expressed by 
\begin{equation}
\overline{P_{et}} = a_{t}\overline{P_{e1}},  
\label{Loykaeq}
\end{equation}
where 
\begin{align}
a_{t} & = \frac{1}{M}\sum_{i=1}^{M}a_{i}, \quad \text{ and } \quad \overline{P_{e1}} \approx \frac{C_{2(N - M)+1}^{N - M + 1}}{(4\gamma_{0})^{N - M + 1}},
\end{align}
with $C_k^n$ denoting the binomial coefficient, and $\gamma_{0}$ the {\it average SNR}. For the sake of space, we refer the reader to~\cite{2006.loyka.vblast.ieee.trans.comm} for the formal definition of the terms $a_{i}$. It is worth noting that the average total BER $\overline{P_{et}}$ is an average with respect to the order of symbol stream detection, as well. In order to verify the accuracy of (\ref{Loykaeq}), we performed a set of Monte Carlo simulations for three antenna configurations. The results are depicted in Figure~\ref{fig:montecarlo}. As we can see, Loyka and Gagnon's analytical model captures V-BLAST performance quite well, and we have adopted their model to implement V-BLAST in the ns-3 simulator. 
Loyka's model takes as input parameter the average before-processing SNR per branch $\gamma_0$, which is assumed to be identical at all branches~\cite{ieeetrans.loyka.gagnon.2004}. 
Thus, for implementation in the simulator, it is computed the average received power across the $N$ {\it receive} antennas, for both the signal of interest (from $i$ to $j$), as well as for the MAI. Hence, the following SINR computation is used as an input to the BER expression provided by Loyka and Gagnon:
\begin{equation}
\mbox{SINR} =  \frac{\frac{1}{N} || \boldsymbol{H}_i^j ||^2_F P_t/M}{P_N + \frac{1}{N} \sum_{k=1}^K ||\boldsymbol{H}_k^j||^2_F P_t/M},
\label{eq:implement sinr}
\end{equation}
where $P_N$ is the average noise power computed over the system bandwidth. 

\subsection{MIMO Channel Simulation}
\label{sec:mimo-channel-simulation}

\begin{figure}[htb]
\centering
 \includegraphics[scale=0.23]{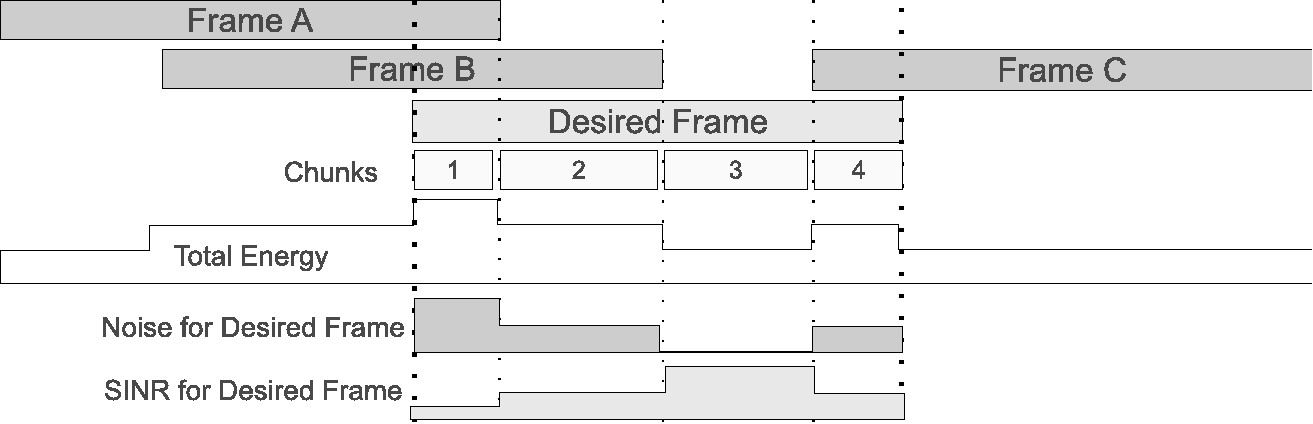}
\caption{Example of chunk segmentation of a frame reception. The received frame (labeled ``desired'') is divided into four unequal parts that have different SINRs due to simultaneous transmissions over the channel of frames A, B and C.}
 \label{fig:chunk}
\end{figure}

Typically, discrete-event network simulators do not allow simulations at the symbol level due to the already high computational cost associated with running a full protocol stack in (potentially) hundreds of nodes. Therefore, the actual mechanisms involved in transmitting/receiving a given signal are normally abstracted through the use of mathematical formulae that relate the signal-to-interference-plus-noise ratio (SINR) with the bit error rate (BER). Hence, in order to know whether a scheduled frame is received without errors, discrete-event simulators typically compute the BER for a given set of bits where both SINR and transmit bit rate are found constant~\cite{lacage.wns2.2006}. This set of bits is called a ``chunk,'' and its contents depend on the frames scheduled for transmission, as depicted in Figure~\ref{fig:chunk}.
For each chunk, the simulator computes the SINR according to
\begin{equation}
\mbox{SINR} = \frac{P_i^j}{P_n+ \sum_k P_k^j},
\label{eq:sinr}
\end{equation}
where $P_i^j$ is the signal power transmitted by node $i$ and received at node $j$, $P_n$ is the average noise power, and $P_k^j$ is the signal power received from each {\it interferer} node $k$ whose transmission overlaps with that particular chunk. Given the SINR, the BER is computed for a specific modulation/coding scheme according to some mathematical formulae. Also, given the system transmission rate, and the chunk's time length, the number of bits $n_b$ contained in the chunk can be calculated. Then, by knowing the BER, the probability $P_{sc}^i$ of successful transmission of the $i$th chunk is computed as $P_{sc}^i = (1 - \text{BER})^{n_b}$, 
where it is assumed independent bit errors within a chunk. Finally, the packet error rate (PER) is computed according to 
\begin{equation}
 \mbox{PER} = 1 - \prod_{i=1}^{n_c}P_{sc}^i,
\end{equation}
where $n_c$ is the number of chunks in the frame.


In order to simulate the effects of a MIMO channel, we extend the simulator's physical layer model to include the numbers $M$ and $N$ of transmit and receive antennas, respectively, as input parameters. Also, the power received at a given node from a {\it specific} transmitter is stored in a matrix $\boldsymbol{P}_{N \times M}$, whose entries $p_{ij}$ represent the power received at antenna $i$ from transmit antenna $j$. Hence, each element $p_{ij}$ is generated by assuming that the transmit power is equally divided among the $M$ transmit antennas (in both Alamouti and V-BLAST systems), and that signal power from each transmit antenna is equally attenuated by the corresponding large-scale path loss model according to the receiver's location in the terrain. In this work, we do not consider shadowing, but it can be easily introduced into the model. Finally, small-scale fading effects are independently introduced at each entry $p_{ij}$ according to the Rayleigh fading model. 


As far as the implementation of channel impairments on frame transmissions is concerned, we follow the approach adopted by many simulators, which is based on applying a {\it single instance} of channel fading (large- and small-scale fading) to {\it all} bits within the frame. In other words, all bits within a frame are attenuated by the same instance of channel impairment. In the MIMO case, however, we associate a {\it single instance} of {\it channel matrix} to the entire frame reception. But, for every {\it transmitted} frame, different channel matrices are instantiated to associate them with {\it every other node in the topology}, according to their location at the time of frame reception. Hence, although we limit the amount of computation required within a frame (by assuming the same channel matrix to all bits in the frame), multiple channel instances are used to associate a given frame with every other node in the topology. Finally, given the received power matrix, one can compute the effective SINR at a given receiver according to the MIMO technique of choice, as explained before. Once the SINR is computed (following the previous ``chunk'' approach), one can proceed with the PER computation for the particular frame. 





\section{Enabling MIMO at the MAC Layer}
\label{sec:mimo_at_mac}

In this section, we present the proposed modfications to a CSMA-like protocol such as the IEEE 802.11 DCF MAC for operation on a MIMO-enabled ad hoc network. Both physical and virtual carrier sensing operations are defined under MIMO transmission. In addition, modifications to the traditional four-way handshake is proposed to accomodate the operation under transmit diversity or spatial multiplexing.

\subsection{Physical Carrier Sense}
\label{sec:physical-carrier-sense}

Apart from signal reception, a key issue that needs to be addressed in a CSMA-like MIMO ad hoc network is how to perform the {\it clear channel assessment} (CCA) operation. This task reflects a node's view of channel availability, and it may have a great impact on the spatial reuse in the network. In the literature, the issue of CCA under MIMO has not been properly investigated. Nonetheless, some works have advocated the use of Alamouti's diversity gains as a means to extend the transmission range of MAC control frames without resorting to extra power or directional antennas~\cite{ieee.levorato.casari.2007, rosseto.trans.wireless.comm.2009}. However, such range extension may incur less spatial reuse. Unfortunately, this side effect has not been properly investigated by previous works because they have only considered fully-connected (i.e., single-hop) networks. 

By using multiple antennas, a receiver captures the energy from $M$ transmit antennas in its $N$ receive antennas. Under a MIMO scheme such as Alamouti or V-BLAST, each stream from each transmit antenna carries $1/M$ of the total transmit energy (assuming the total transmit energy is equally divided among all $M$ antennas). Therefore, a receiver will detect the {\it total} transmit energy in {\it each} receive antenna (under path loss and multipath fading). Consequently, the {\it total} received energy is, roughly speaking, $N$ times the energy received on a single antenna. Therefore, depending on how the total received energy is handled, the perception of channel activity may vary considerably. In this paper, two CCA mechanisms are investigated: the first one is based on the {\it total received energy} (or ``sum energy'') detected by all antennas. In practice, this corresponds to detecting an {\it aggregate} received power (collected from all antennas) above a given threshold. The second CCA mechanism is based on the {\it average energy} received across all antennas. To compute that, the aggregate power received  at all antennas is divided by the number $N$ of receive antennas. This average aggregate power is then compared to a given threshold in order to determine if the channel is idle or busy. 

\subsection{Virtual Carrier Sense}
\label{sec:virtual-carrier-sense}

In addition to physical carrier sense, the IEEE 802.11 DCF MAC specifies a {\it duration field} in the header of RTS, CTS, DATA, and ACK frames 
in order to inform neighboring nodes about the total time the medium will be busy for the impending communication. Neighboring nodes who receive this information update their network allocation vector (NAV) accordingly. Hence, nodes perform both physical and virtual carrier sense (i.e., check their NAV) before attempting a four-way handshake with someone. The information in the duration field considers both the number of bits and the transmission rate at which frames will be transmitted. 

Compared to a SISO configuration, V-BLAST delivers an M-fold gain in frame transmission time because it uses $M >$ 1 antennas for parallel transmission of $M$ consecutive symbols at {\it each} symbol time. Therefore, if V-BLAST is adopted, the number $M$ of transmit antennas must be taken into account in order to set appropriate values in the {\it duration field} of MAC headers of control and DATA frames. Differently from V-BLAST, however, the Alamouti scheme does not reduce the frame transmission time if compared to a SISO scheme. This is because, in the end, its operation consumes a number of symbol times equivalent to the number of symbols to be transmitted. Therefore, under Alamouti, the time at which the medium is kept busy  does not change with respect to a SISO scheme. Given these considerations, if both MIMO schemes are to be supported by the underlying PHY layer, the MAC sublayer will need to know not only the number of bits and transmission rate in order to properly set the MAC header's duration field, but also the number $M$ of transmit antennas and the MIMO technique of choice. 

\begin{figure}[htb]
 \centering
 \includegraphics[width=0.4\textwidth,bb=0 0 1009 778]{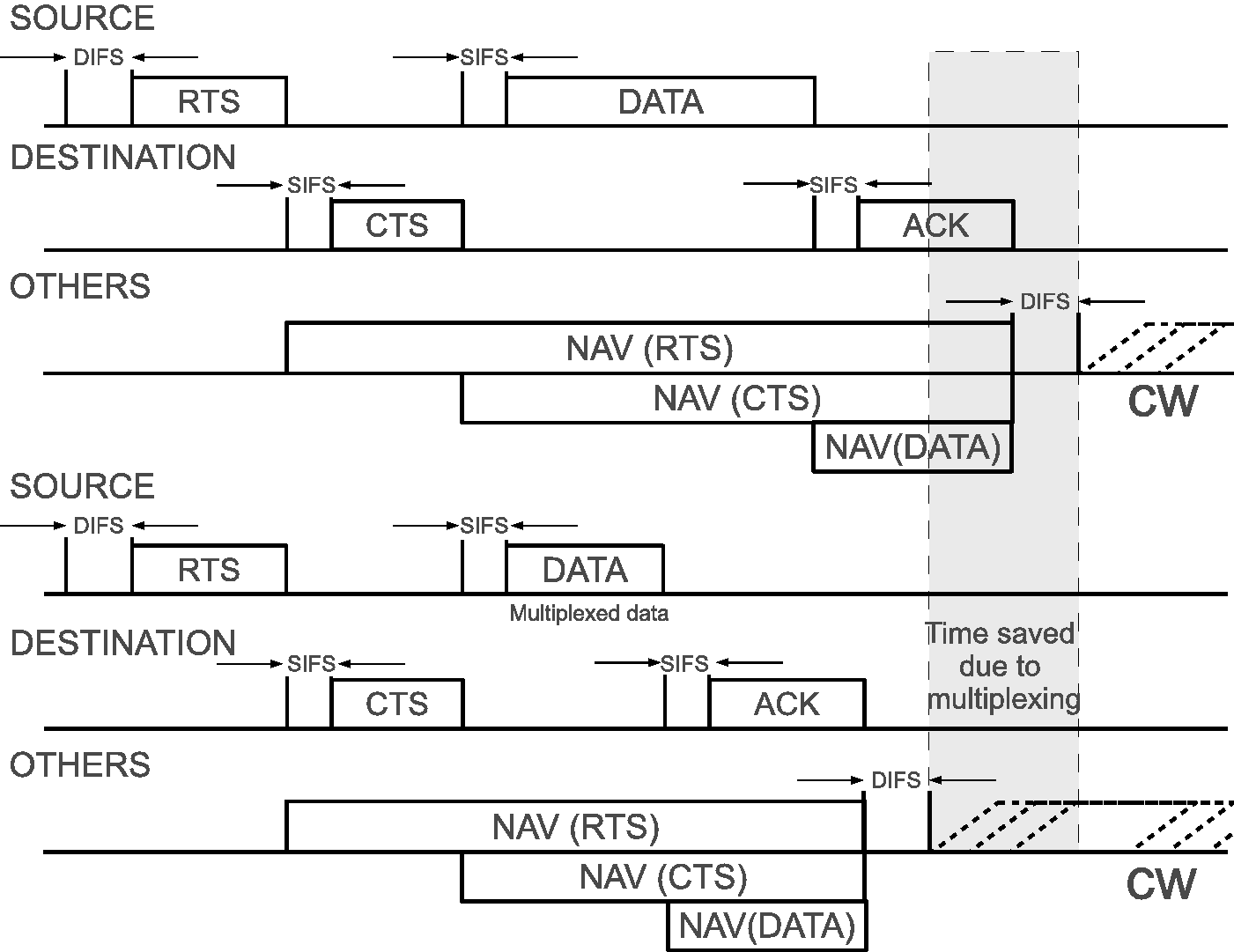}
 \caption{Comparison of the four-way handshake duration between SISO (top) and MIMO spatial multiplexing (bottom). The gray area shows the relative time savings due to spatial multiplexing applied to the DATA frame only.} 
 \label{fig:rtscts}
\end{figure}

Figure~\ref{fig:rtscts} depicts a comparison between two four-way handshakes: at the top, it shows a four-way handshake under SISO mode, whereas the diagram in the bottom shows a DATA frame transmission under V-BLAST with two transmit antennas. With V-BLAST, the DATA transmission time is reduced by half, and the channel becomes idle sooner, which allows other nodes to have access to the channel earlier. Notice that, since the DATA frame is the only frame whose transmission is carried out under MIMO, it is clear that, the higher the number of transmit antennas, the higher the impact of V-BLAST on the reduction of the four-way handshake time, especially if the DATA frame is long. In Section~\ref{sec:mac-enhancements} we discuss the needed modifications to the MAC layer to allow the {\it joint} operation of Alamouti and V-BLAST at each node.   


\section{Performance Evaluation with Individual MIMO Schemes}
\label{sec:individual_mimo_schemes}

This section presents the general setup for the simulations carried out in this work, along with a performance evaluation of ad hoc networks where all nodes utilize the same MIMO scheme under different antenna configurations. Before that, the choice of appropriate clear channel assessment scheme under MIMO is also investigated. The purpose of these studies is to understand the strengths and limitations of each MIMO scheme, so we can propose an appropriate joint operation scheme in Section~\ref{sec:joint-mimo}. 




\subsection{Performance Evaluation Scenarios}
\label{sec:sim-scenarios}

\begin{figure}[hbt]
\centering
\begin{tabular}{c}
 \includegraphics[scale=0.34]{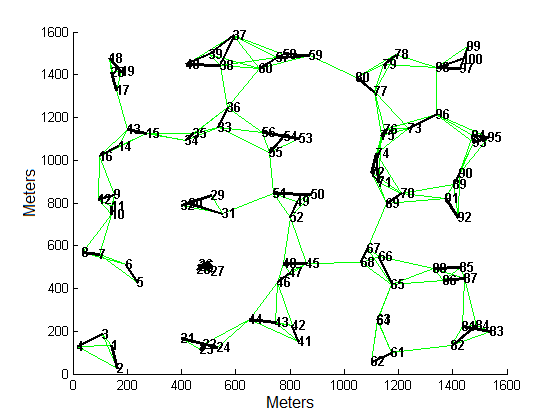} \\
 {\scriptsize (a)} \\
 \includegraphics[scale=0.34]{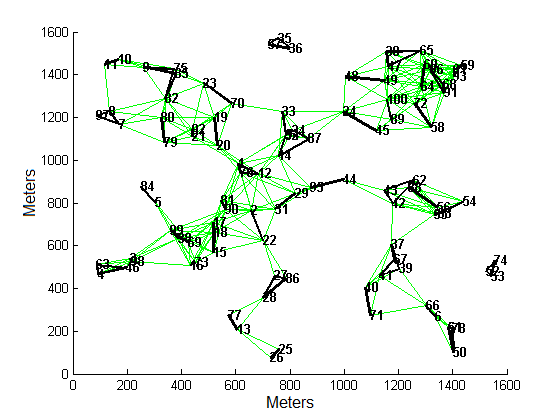} \\
 {\scriptsize (b)} \\
 \includegraphics[scale=0.34]{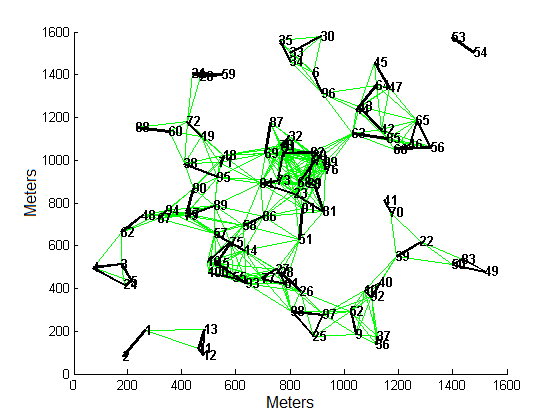} \\
 {\scriptsize (c)}
\end{tabular}
\caption{Examples of topologies used in simulations with different contention levels. The numbers indicate the node, and the dark lines indicate source-destination pairs. The light lines indicate nodes within the sensing range.}
\label{fig:topologias}
\end{figure}

All performance evaluations are carried out by taking into account three figures of merit, all at the MAC level: the {\it average throughput per node}, the {\it average delay per frame}, and {\it fairness in throughput}. By MAC-level throughput one must understand it as the data throughput achieved by the MAC protocol at each link between a pair of neighboring nodes. In other words, the end-to-end throghput over a path, as a result of routing activities, is not a concern in this paper. The focus is on {\it static} scenarios under {\it saturated} traffic, so that the impact of high contention on average throughput is investigated.  Hence, the application layer at each node generates packets frequently enough so that a DATA frame is always ready to be sent at the head of the transmit queue. At the same time, the node itself is the target receiver of some other neighboring node(s) (i.e., all nodes act as both transmitters and receivers). The average DATA frame delay is computed between the instant when the sender's IP layer conveys a data packet to the MAC transmit queue, and the instant when the frame is successfully received at the destination's IP layer. For fairness computation, the popular Jain's fairness index is adopted. Nine topologies are used with 100 nodes placed on a 1600~$\times$~1600~m terrain. Transmitter/receiver pairs are distant apart by a maximum transmission range of 150~m under SISO transmission and the two-ray path loss propagation model only (i.e., when we measure it without considering the effects of random small-scale fading). 
Each simulation run corresponds to 60 seconds of CBR traffic over UDP at each node, in each topology. Table~\ref{tab:param} summarizes the rest of PHY- and MAC-layer parameters used in simulations.
\begin{table}[htb]
 \centering
 \caption{PHY- and MAC-Layer Parameters}
{\small \begin{tabular}{ll} \hline 
 Energy Detection Threshold & -73.8764 dBm \\
 Clear Channel Threshold & -80.9201 dBm \\
 Noise Figure & 7 dB \\
 Transmit Power & 10 dBm \\
 Antenna Height & 1.2 m \\
 Data Packet Size & 1412 bytes \\
 DSSS PHY Trasmission Rate & 1 Mbps \\
 Modulation & DBPSK \\
 Transmit queue size & 400 \\ \hline 
\end{tabular}
}
\label{tab:param}
\end{table}
The clear channel assessment (CCA) threshold is set to allow a carrier sensing range of 225~m with a SISO antenna configuration (the impact of multiple antennas on the sensing range will be presented shortly). 
Under such constraints, topologies {\it are not} fully connected, and concurrent transmissions may occur due to spatial reuse. In addition, the performance under different contention levels is investigated by modifying the {\it sparsity} of nodes in the terrain. By doing that, we vary the average number of nodes that can be sensed by any particular node. Hence, the selected network topologies are loosely divided into three general groups (with three topologies in each group): {\it low-contention} topologies, where nodes are sparsely distributed over the terrain, {\it medium-contention} topologies, which have a higher degree of connection among neighboring nodes, and {\it high-contention} topologies. Figure~\ref{fig:topologias} displays sample topologies from each contention group.

\subsection{Clear Channel Assessment under MIMO}
\label{sec:impact-multiple-antennas}

\begin{figure}[htb]
   \centering
   \includegraphics[scale=0.4]{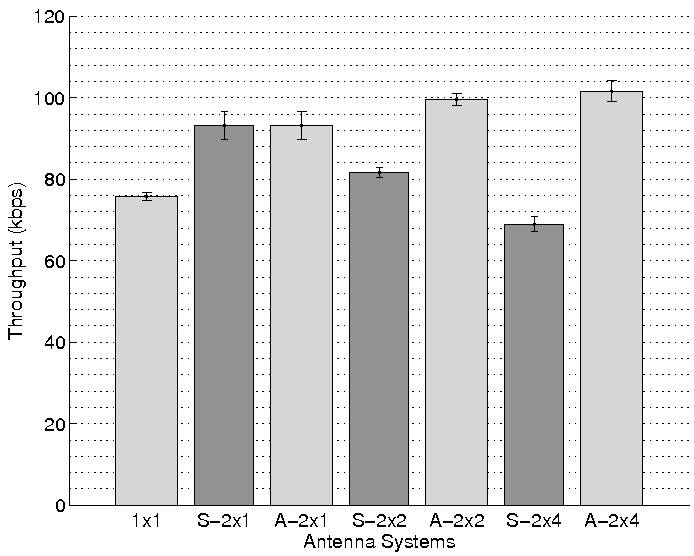}
   \centering
   \caption{Average network throughput versus CCA method under $M \times N$ antennas. A-$M \times N$ indicates average received energy, and S-$M \times N$ indicates the sum energy.}
   \label{fig:cca-computation-comparison}
\end{figure}

In this section, we investigate the issue of performing CCA under multiple antennas according to the mechanisms described in Section~\ref{sec:physical-carrier-sense}. In particular, we study the impact of using either the ``sum energy'' or the ``average energy'' as the CCA method of choice under the Alamouti Scheme, since this scheme allows range extension without increasing signal power. Figure~\ref{fig:cca-computation-comparison} depicts the impact of the CCA method on the average throughput per node under Rayleigh fading. The average results are computed over all nodes and topologies discussed previously. For purposes of comparison, we also include the results for a SISO IEEE 802.11 network. As we can see, if CCA is based on the {\it sum energy} (indicated by `S-$M \times N$' in the graph), average throughput decreases because each node starts sensing further distant nodes as $N > 1$. This leads to higher backoff activity and, consequently, throughput degradation due to less spatial reuse in the network. Therefore, in this case, Alamouti's diversity gains are {\it cancelled out} by the carrier sense activity of nodes. On the other hand, if CCA uses the average energy received across all antennas (indicated by `A-$M \times N$' in the graph), higher spatial reuse is achieved while, at the same time, link-level retransmissions become less frequent as more receive antennas are added (due to Alamouti's diversity gains). Consequently, the average throughput per node increases as the number of receive antennas increases. 
Therefore, one must be careful when seeking range extension of control frames through the use of diversity schemes in ad hoc networks: the added range extension may result in poor spatial reuse. This, in turn, may lead to overall throughput degradation. Such a problem has not been properly addressed in the literature before. Based on these results, in the following, we adopt the {\it average energy received across all antennas} as the CCA method of choice for operation of all MIMO ad hoc networks in this work.

\subsection{MIMO Ad Hoc Networks with Alamouti Scheme}
\label{sec:performance_alamouti}

\begin{figure}[htb]
 \centering
 \includegraphics[scale=0.4]{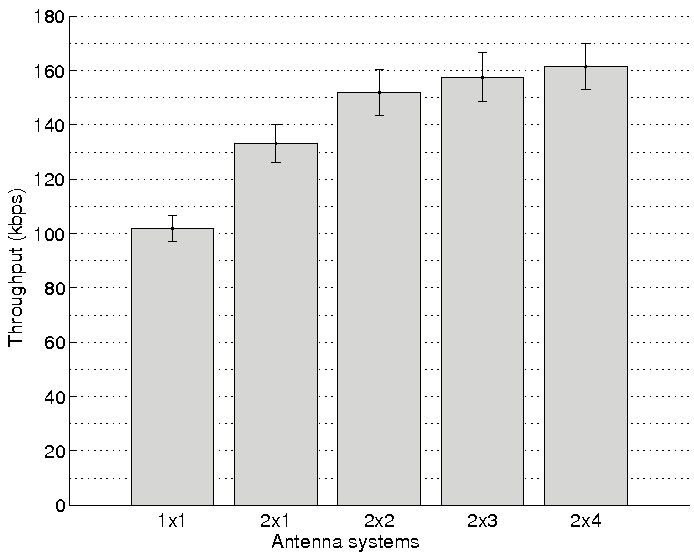}
 \caption{Average throughput per node for different antenna configurations under the Alamouti scheme.}
 \label{fig:alamthrp}
\end{figure}

This section presents the simulation results for ad hoc networks equipped with the Alamouti scheme according to the mechanisms described in Section~\ref{sec:alamouti_scheme}. Figure~\ref{fig:alamthrp} depicts the average throughput per node under Rayleigh fading for different antenna configurations ($M \text{ transmit} \times N$ receive antennas). Error bars indicate confidence intervals of 95\%.  As expected, throughput performance improves as the number of receive antennas increases. This is due to Alamouti's well-known diversity gains, which improve robustness to channel errors as more receive antennas are added. Considering the investigated scenarios, the gains in throughput with respect to a SISO IEEE 802.11 network are about 30.8\% (2$\times$1), 49.2\% (2$\times$2), 54.8\% (2$\times$3), and 58.6\% (2$\times$4), respectively. It can be observed that throughput gains approach saturation as the number of receive antennas increases. This is because bit error rates at each link tend to achieve their best performance due to diversity gains, and adding more receive antennas has only incremental impact on bit error rate. From that point on, throughput becomes fundamentally limited by MAC-level contention and underlying mechanisms of the MAC protocol in place. In addition, the attainable gains are highly dependent on the underlying carrier sensing range, which is controlled by both CCA threshold and how CCA is performed with multiple antennas. 

\begin{figure}[htb]
 \centering
 \includegraphics[scale=0.4]{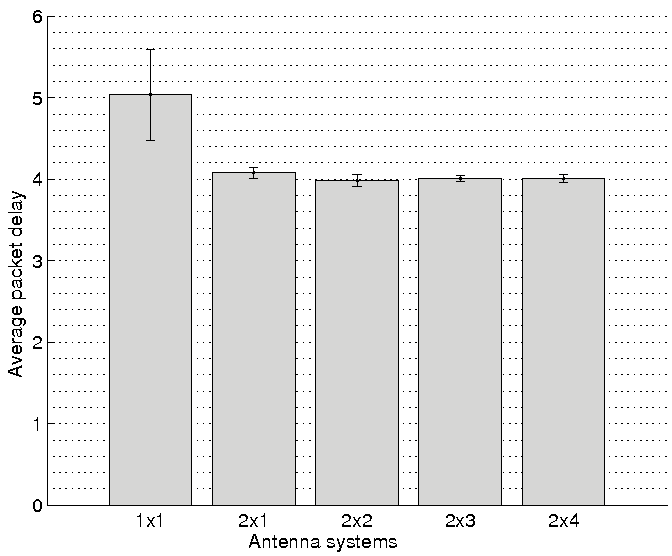}
 \caption{Average point-to-point delay for different antenna configurations under the Alamouti scheme.}
 \label{fig:alamdlay}
\end{figure}

Figure~\ref{fig:alamdlay} contains the results for the average point-to-point delay per node. Considering that traffic is saturated at each node (i.e., the application generates packets at a rate much higher than what can be delivered by lower layers), a significant decrease in delay (about 1.0~s) occurs when Alamouti is first used (the 2$\times$1 case). Once robustness to channel errors improves, less frames are retransmitted by the MAC layer. At the same time, we observe that delay does not change significantly as the number of receive antennas increases. This is because not only the incremental gains in throughput start decreasing, but the resulting point-to-point delay is dominated by channel contention, the mechanisms of the BEB algorithm, and the waiting time at the transmit queue. 

\begin{figure}[htb]
 \centering
 \includegraphics[scale=0.4]{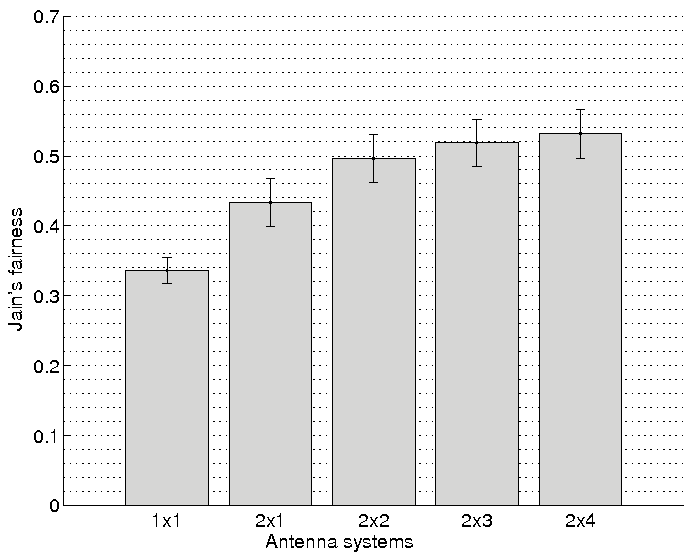}
 \caption{Jain's fairness index for different antenna configurations under the Alamouti scheme.}
 \label{fig:alamfair}
\end{figure}

As far as fairness is concerned, Figure~\ref{fig:alamfair} shows that improvements in fairness follow the gains in throughput. Such a result indicates that robustness to channel errors allow nodes to spend less time in frame retransmissions, which make them less susceptible to the fairness issues associated to the BEB algorithm of the IEEE 802.11 DCF: the nodes who last acquired the channel are the ones more likely to acquire it again. 
It is interesting to observe how poor the fairness of a SISO network is in the scenarios investigated, even though spatial reuse and low-contention scenarios are considered. With the addition of the Alamouti scheme, fairness improves by 51\% when using the 2$\times$2 configuration, por example.

\subsection{MIMO Ad Hoc Networks with V-BLAST}
\label{sec:desempenho_vblast}

\begin{figure}[htb]
 \centering
 \includegraphics[scale=0.4]{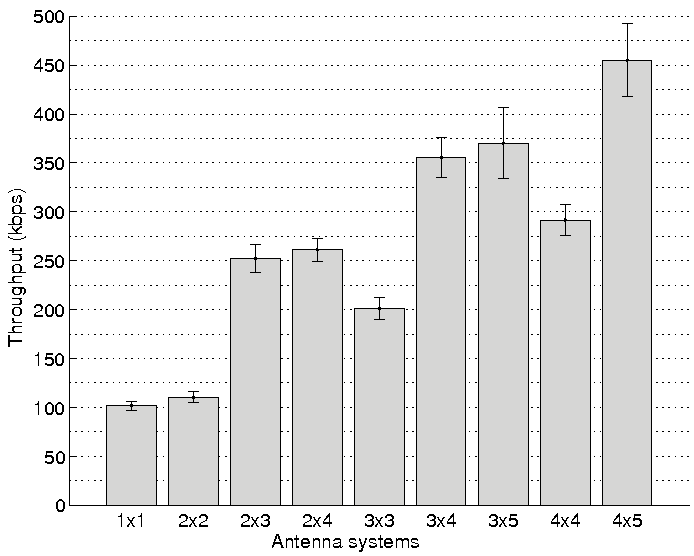}
 \caption{Average network throughput with only V-BLAST system enabled, under Rayleigh fading ($K=0$) for different antenna configurations.}
 \label{fig:smuxthrp}
\end{figure}

The simulation results for ad hoc networks equipped with the V-BLAST system are presented in this section. Figure~\ref{fig:smuxthrp} shows the average throughput per node under different antenna configurations.
According to capacity results of MIMO systems~\cite{goldsmith}, one should expect a linear increase in data rate when more antennas are added for transmission and/or reception. However, the results show that this linear increase is not observed at all cases. The reason for this fact is better explained through the concept of diversity. Loyka and Gagnon~\cite{ieeetrans.loyka.gagnon.2004} have shown that the {\it diversity order} of V-BLAST at the $i\text{-th}$ processing step is $N - M + i$ (under Rayleigh fading and optimal ordering). Such a result can be readily verified in Figure~\ref{fig:montecarlo}, which reflects the average diversity gain across all processing steps. Thus, for a fixed number $M$ of transmit antennas, the effect of increasing the number $N$ of receive antennas translates into a change in the slope of the curves, i.e., for a given SNR, the BER decreases as the number of receive antennas increases (observe the cases $2\times 2$ and $2\times 3$). 

As already mentioned, the effect of diversity is to improve transmission reliability, so that we can expect less time spent on retransmissions. Hence, by considering only the cases when $M < N$ (i.e., configurations 2$\times$3, 3$\times$4, and 4$\times$5), one can observe a {\it quasi-linear} increase in average throughput with respect to min$\{N,M\}$. More specifically, the average throughput for the 2$\times$3 system (266.73 kb/s) is more than twice (2.62) the SISO case (101.78 kb/s), while the 3$\times$4 configuration delivers an average throughput of 355.58 kb/s, which is about 3.49 times higher than SISO, and the 4$\times$5 configuration achieves a gain of 4.47 with an average throughput of 455.11~kb/s. However, in spite of the significant throughput gains obtained when $M < N$, no further significant improvement is perceived if more receive antennas are added while keeping the number of transmit antennas fixed (e.g., switching from 3$\times$4 to 3$\times$5 in Figure~\ref{fig:smuxthrp}). This is because no matter how small the BER becomes as $N$ increases (for a given SINR), there is no additional data multiplexing on the transmitter (since $M$ is kept fixed). Consequently, one should expect the average throughput to achieve a saturation if only the number of receive antennas increases. 

On the other hand, it is interesting to observe the diversity-multiplexing trade-off when the number of receive antennas is kept fixed while the number of transmit antennas increases. From the results, one may conclude that it is better to guarantee robust transmissions with less data multiplexing (e.g., the 3$\times$4 case) than to add one more antenna to speed up transmissions and release the channel earlier (the 4$\times$4 case). Here, it is important to emphasize that, although these gains in robustness are expected from V-BLAST at a single link (see Figure~\ref{fig:montecarlo}), such a result cannot be easily translated into higher-layer performance improvements, since it will depend on source-destination distances (dictated by the underlying topology), and corresponding channel contention (dictated by the MAC protocol) which, if properly designed, could take better advantage of the earlier release of the channel.  

\begin{figure}[htb]
 \centering
 \includegraphics[scale=0.4]{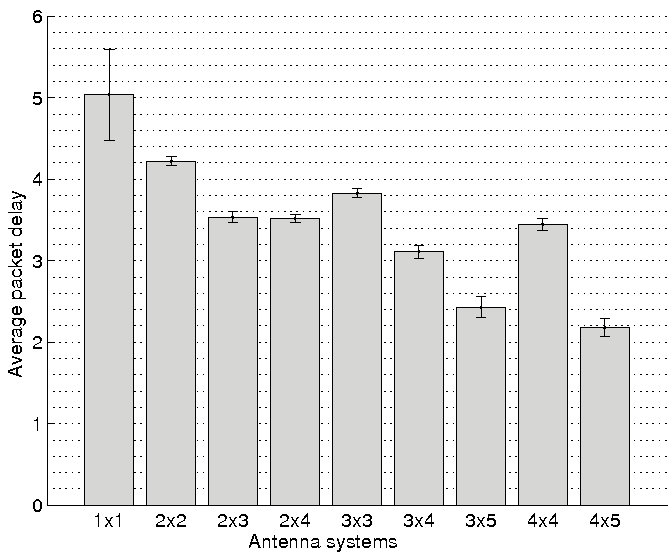}
 \caption{Average point-to-point delay for V-BLAST under Rayleigh fading for different antenna configurations.}
 \label{fig:smuxdlay}
\end{figure}

Figure~\ref{fig:smuxdlay} displays the results for the average point-to-point delay. It can be noticed that the gains with respect to SISO are not in the same order of magnitude of throughput. This is because delay is dominated by waiting times in queues (queue size is 400 frames). The 2$\times$2 configuration provides a reduction in delay that is worse than its counterpart in Alamouti. In this case, the delay reduces by only 11.8\% with respect to SISO. The configurations 2$\times$3 and 2$\times$4 do perform better, as expected, but provide similar gains in delay, of about 43\%. But, when the number of transmit antenas changes to 3, a gradual reduction in delay can be observed as more receive antennas are added. The gains in delay when switching from 3$\times$3 to 3$\times$4 is about 18.8\%, and from 3$\times$3 to 3$\times$5 is about 58.3\%, which means an overall gain of 108\% over SISO. As before, using a 4$\times$4 system is worse than using a 3$\times$4 configuration, on average. However, the addition of one more receive antenna (3$\times$5) delivers the best gain over SISO, which is about 127.3\%. 

\begin{figure}[htb]
 \centering
 \includegraphics[scale=0.4]{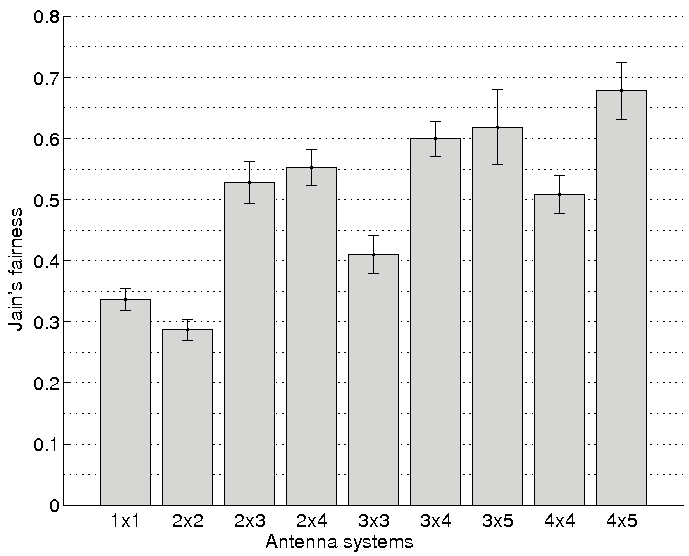}
 \caption{Jain's fairness index for V-BLAST under Rayleigh fading for different antenna configurations.}
 \label{fig:smuxfair}
\end{figure}

Regarding fairness, Figure~\ref{fig:smuxfair} depicts the resuts for Jain's fairness index. It can be noticed that fairness improves significantly with spatial multiplexing, especially under $M < N$ configurations. Again, such a result indicates that, the sooner the nodes finish their transmissions, the highest the odds that the channel is acquired more evenly by competing nodes. Fairness improves by 77\% with a 3$\times$4 configuration (over SISO), and about 100\% with a 4$\times$5 system. 

\begin{figure}[htb]
\centering
 \includegraphics[scale=0.4]{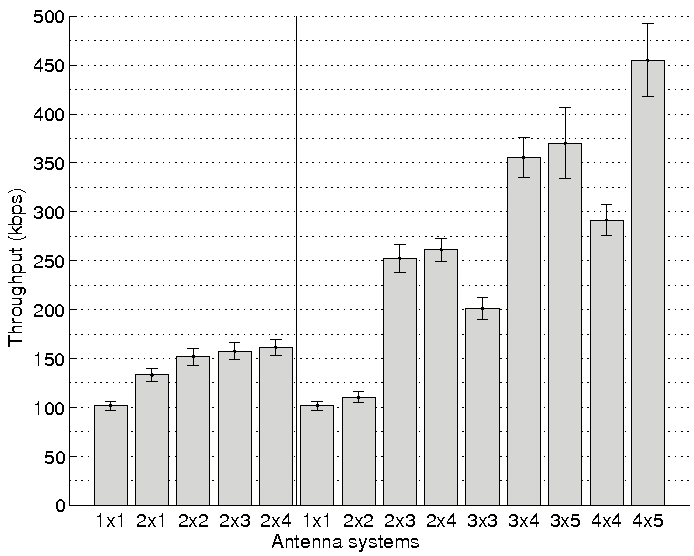}
 \caption{Average throughput per node for the Alamouti scheme, on the left hand side of the vertical line, and V-BLAST, on the right hand side.}
\label{fig:alamsmuxthrp}
\end{figure}

Finally, Figure~\ref{fig:alamsmuxthrp} displays a comparison between V-BLAST and the Alamouti scheme regarding the average throughput per node. Alamouti's performance is depicted on the left hand side of the vertical line, while the performance of V-BLAST is shown on the right hand side. For purposes of comparison, it also shows the results of V-BLAST for the cases $M = 3$ and $M = 4$. 
Considering only the case $M = 2$, V-BLAST provides the best results if $M < N$. In such cases, V-BLAST not only implements spatial multiplexing, but also allows diversity gains. On the other hand, Alamouti outperforms V-BLAST in the 2$\times$2 case.

	\section{Joint MIMO Operation} 
\label{sec:joint-mimo}

The study of ad hoc networks whose nodes are all equipped with the same MIMO scheme allows us to understand the performance of each technique in the scenarios of interest. 
In spite of the generally higher average performance delivered by V-BLAST, it is reasonable to expect that a fraction of the nodes may experience very low link quality due to large- and small-scale path loss propagation effects, and other signal impairments. For such nodes, an error-resilient link is at a premium, and the Alamouti scheme might work as the best choice. On the other hand, if the perceived SINR is very high, one would rather use all antennas for transmission and reception (i.e., $M = N$) to achieve higher data rates under V-BLAST. Consequently, it is reasonable to expect that average network performance might improve if each pair of nodes could select the most appropriate MIMO scheme for their communication. In fact, adaptive MIMO schemes have been proposed before, but mostly within the context of link optimization only, i.e., decisions on ``which MIMO scheme'' and ``when to switch'' are generally tied to link conditions only, without taking into account overall network conditions. In this work, we aim at making a pair of nodes to switch to a given MIMO scheme as a means to improve overall network performance. 
In the following, we present the needed modifications to the MAC protocol discussed in Section~\ref{sec:mimo_at_mac} to allow the joint operation of more than one MIMO scheme.


\subsection{MAC Enhancements for Joint MIMO Operation}
\label{sec:mac-enhancements}



The main ideia for the joint operation of multiple MIMO schemes is to let the {\it intended receiver} of a DATA frame to select the best MIMO configuration for the impending communication. This is because the intended recipient of a DATA frame is the one that is (obviously) better positioned to evaluate channel conditions for a sucessful DATA frame reception. To accomplish that, the PHY layer at the intended receiver must compute the {\it average} SINR during reception of an RTS. In our scheme, all control frames are transmitted in SISO mode, except for the DATA frame, which is transmitted with the chosen MIMO scheme (this is to allow the correct reception of control frames by all surrounding nodes). Hence, the RTS {\it time duration} field announces a conservative channel occupation time beforehand: it assumes that no multiplexing will occur, and therefore, time duration is announced based on a SISO transmission of the DATA frame. The average SINR value $\overline{\text{SINR}}$ is passed on to the MAC sublayer, which selects the most appropriate MIMO scheme for DATA frame reception. Based on previous results, and to avoid too much complexity, we propose the use of only {\it three} modes of operation. In particular, if we assume that each node is equipped with $N \ge 3$ antennas, and that all the $N$ antennas are used for reception, the following modes of operation are proposed:
\begin{enumerate}
\item[1\textordmasculine )] {\it Full Transmit Diversity:} 
   \begin{itemize}
   \item Alamouti Scheme with $2 \times N$ configuration; 
   \end{itemize}
\item[2\textordmasculine )] {\it Spatial Multiplexing with Diversity:} 
   \begin{itemize}
   \item V-BLAST with $M = N - 1$ configuration;
   \end{itemize}
\item[3\textordmasculine )] {\it Full Spatial Multiplexing:} 
   \begin{itemize}
   \item V-BLAST with $M = N$ configuration.
   \end{itemize} 
\end{enumerate}

The intended receiver chooses the MIMO scheme by comparing the $\overline{\text{SINR}}$ value with two {\it pre-defined} thresholds: $\text{SINR}_{\min}$ and $\text{SINR}_{\max}$. More specifically, the following algorithm is used for MIMO mode selection at the receiver: \medskip 

\indent {\bf if} $\overline{\text{SINR}} < \text{SINR}_{\min}$, \medskip

\hspace{1cm}  Choose {\it Full Transmit Diversity} \medskip

\indent {\bf else if} $\text{SINR}_{\min} \le \overline{\text{SINR}} < \text{SINR}_{\max}$, \medskip

\hspace{1cm} Choose {\it Spatial Multiplexing with Diversity}  \medskip

\indent \hspace{0.6cm} {\bf else} 

\hspace{1cm} Choose {\it Full Spatial Multiplexing}. \\ 

The values $\text{SINR}_{\min}$ and $\text{SINR}_{\max}$ are empirically found based on {\it average network performance}, according to an experimental procedure to be detailed shortly. The chosen MIMO scheme is conveyed back to the sender of the RTS through an extra byte that is added to the CTS control frame (this header field has the length of a byte in case one wants to define other MIMO schemes in the future). 
Once the MIMO scheme has been chosen, the {\it time duration field} of the CTS must be set. 
As a result, all neighbors of the intended receiver will have their {\it network allocation vector} (NAV) updated correctly regarding the time duration of the upcoming data transfer. When the CTS frame reaches the sender of the RTS, the information regarding the MIMO scheme is read from the CTS header and passed on to the PHY layer in order to switch to the appropriate MIMO configuration. Also, the time duration field of the DATA frame is updated accordingly, so that the sender's neighbors are updated about that, too. 
Next, we detail the experimental procedure used to find appropriate values for $\text{SINR}_{\min}$ and $\text{SINR}_{\max}$. While it is clear that the proposed threshold values may not work in all possible network scenarios, the goal of this approach is to show how these parameters can be tuned to specific application scenarios and network conditions.

\subsection{Setting the Lower Threshold $\text{SINR}_{\min}$} 
\label{sec:hyba}

\begin{figure}[htb]
 \centering
 \includegraphics[width=0.4\textwidth]{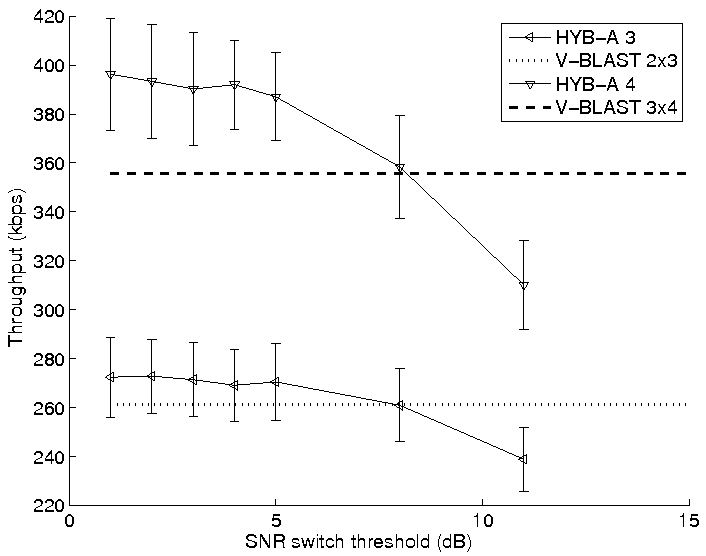}
 \caption{Average network throughput when operating in HYB-A mode as a funcion of the switching inferior threshold with 3 and 4 antennas available for transmission in comparisson to the average throughput of the pure V-BLAST($M=N-1$) system, indicated by horizontal lines.}
 \label{fig:hyba3hyba4thrp}
\end{figure}

In this experiment, we want to find the value of $\text{SINR}_{\min}$ that maximizes network performance if nodes are allowed to switch only between Alamouti ($2 \times N$) and V-BLAST ($M = N - 1$). For that, we set a value of $\text{SINR}_{\max}$ high enough so that nodes {\it never} switch to V-BLAST ($N = M$). We name this mode of operation ``HYB-A'' (from ``hybrid mode'' A). Hence, we vary the values of $\text{SINR}_{\min}$ within a pre-defined set that ranges from 2 to 15 dB and, for each value of $\text{SINR}_{\min}$, we compute the average throughput per node across all topology scenarios described in Section~\ref{sec:sim-scenarios}. Figure~\ref{fig:hyba3hyba4thrp} depicts the results for the average throughput per node for both 2$\times$3 and 3$\times$4 cases, according to different values for $\text{SINR}_{\min}$. These results are indicated as ``HYB-A 3'' and ``HYB-A 4,'' respectively, based on the number $N$ of receive antennas. All results correspond to a confidence level of 95\%. For purposes of comparison, we have also included a horizontal line in the graph that indicates the average performance when nodes use only V-BLAST with 2$\times$3 and 3$\times$4 configurations, respectively.

As we can see, the average throughput per node decreases as the value of $\text{SINR}_{\min}$ increases. This is true for both $N = 3$ and $N = 4$. As $\text{SINR}_{\min}$ increases, more pairs of nodes switch to the Alamouti scheme, and performance starts deteriorating, since Alamouti's performance is generally lower than V-BLAST for such antenna configurations (see Figure~\ref{fig:alamsmuxthrp}). But, when $\text{SINR}_{\min} <$~8 dB, HYB-A provides performance gains over pure V-BLAST. This means that pairs of nodes that only use V-BLAST ($M = N - 1$) have higher frame discard rates compared to the Alamouti scheme when $\overline{\text{SINR}}<$~8~dB. Hence, as frame discards decrease, the average network throughput increases. The best performance gains observed were 8.11\% ($N = 3$) and 10.24\% ($N = 4$). 


\subsection{Setting the Upper Threshold $\text{SINR}_{\max}$} \label{sec:hybb}

\begin{figure}[htb]
 \centering
 \includegraphics[width=0.4\textwidth]{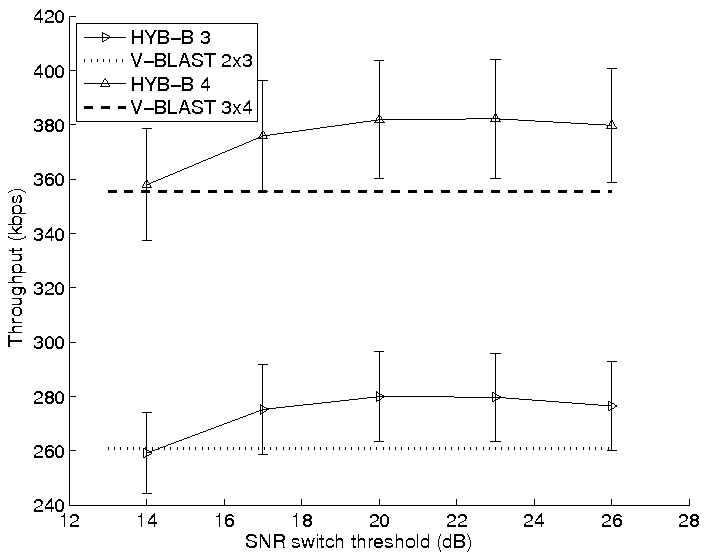}
 \caption{Average network throughput when operating in HYB-B mode as a funcion of the switching inferior threshold with 3 and 4 antennas available for transmission in comparisson to the average throughput of the pure V-BLAST($M=N-1$) system, indicated by horizontal lines.}
 \label{fig:hybb3hybb4thrp}
\end{figure}

In this study, we want to find the value of $\text{SINR}_{\max}$ that maximizes network performance when nodes are only allowed to switch between V-BLAST ($M=N-1$) and V-BLAST ($M=N$). For that, we set the value of $\text{SINR}_{\min}$ low enough so that nodes never switch to the Alamouti scheme. We name this mode of operation as ``HYB-B'' (from ``hybrid mode B''). Hence, the values of $\text{SINR}_{\max}$ are varied over a pre-defined set that ranges from 14 to 26 dB. Figure~\ref{fig:hybb3hybb4thrp} shows the average throughput per node across all topology scenarios for both 2$\times$3 and 3$\times$4 cases. Again, for purposes of comparison, we put a line in the graph that indicates the average throughput performance when nodes use only V-BLAST ($M = N - 1$). 
It can be noticed that throughput {\it decreases} as $\text{SINR}_{\max}$ assumes lower values, both for $N = 3$ and $N = 4$. In such cases (lower $\text{SINR}_{\max}$ values), more pairs of nodes switch to full multiplexing ($M = N$) under not-so-high $\overline{\text{SINR}}$ values. Since bit error rates for V-BLAST ($M = N$) are generally higher than V-BLAST ($M = N - 1$), average throughput {\it decreases} as $\text{SINR}_{\max}$ decreases. On the other hand, as $\text{SINR}_{\max}$ increases, only the pairs of nodes that experience higher $\overline{\text{SINR}}$ values switch to V-BLAST ($M = N$). Because of that, these pairs take advantage of full multiplexing under relatively low bit error rates. As a result, average throghput performance increases. The best performance gains are achieved when $20 \le \text{SINR}_{\max} \le 23$~dB, in which case the gains with respect to pure V-BLAST ($M=N-1$) are about 10.96\% ($N = 3$) and 7.51\% ($N = 4$).
 

\subsection{Joint MIMO Operation} 
\label{sec:joint-mimo-operation}

According to previous results, it is possible to obtain some performance gain if  nodes are allowed to switch between the MIMO modes of operation defined in Section~\ref{sec:mac-enhancements}, based on the $\overline{\text{SINR}}$ value computed during RTS reception. Hence, in this section, we investigate the attainable performance gains when nodes operate according to the MAC enhancements presented in Section~\ref{sec:mac-enhancements}. For that, we set $\text{SINR}_{\min}=5$~dB and $\text{SINR}_{\max} = 23$~dB. These are the values that appear to yield the best performance gains under relatively high $\text{SINR}_{\min}$ and low $\text{SINR}_{\max}$, respectively. Under such circumstances, more pairs of nodes are allowed to switch among the three possible MIMO modes. For purposes of comparison, the joint MIMO operation is named ``HYB-C'' (from ``hybrid mode C''). 

\begin{figure}[htb]
 \centering
 \includegraphics[width=0.4\textwidth]{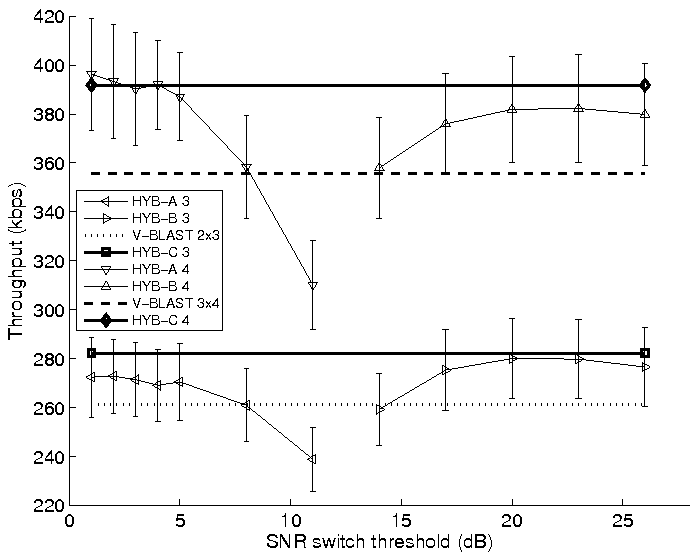}
 \caption{Average network throughput when operating in HYB-C mode (bold line) in comparisson to the HYB-A and HYB-B modes and the V-BLAST($M=N-1$) system for configurations of 3 and 4 antennas.}
 \label{fig:hyb3hyb4thrp}
\end{figure}

Figure~\ref{fig:hyb3hyb4thrp} contains a comparison between the average throughput per node achieved under HYB-C (with the chosen values of $\text{SINR}_{\min}$ and $\text{SINR}_{\max}$), and the performance of pure V-BLAST ($M = N - 1$), HYB-A, and HYB-B (these last two under different values of SINR thresholds).
It can be noticed that HYB-C (bold line) performs better than pure V-BLAST (dashed line), HYB-A, and HYB-B under practically all values of SINR thresholds, and for both $N = 3$ and $N = 4$. The gains with respect to pure V-BLAST are 11.84\% for $N = 3$, and 10.19\% for $N = 4$. 

\begin{figure}[htb]
 \centering
 \includegraphics[scale=0.45]{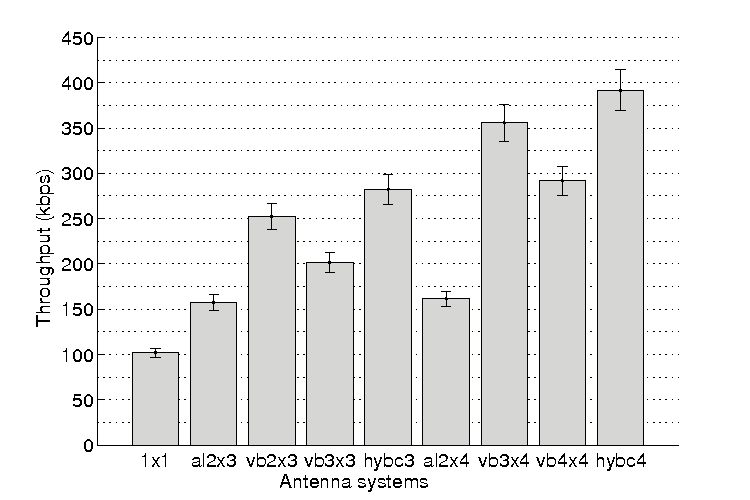}
 \caption{Average throughput per node for different modes of operation under different antenna configurations. The sufix ``al'' indicates the Alamouti scheme, ``vb'', the V-BLAST system, and ``hybc'', the HYB-C operation mode.}
 \label{fig:c_hybcthrp}
\end{figure}

Figure~\ref{fig:c_hybcthrp} shows a performance comparison in terms of {\it average throughput per node} between HYB-C, pure V-BLAST ($M = N - 1$ and $M = N$), Alamouti ($2 \times N$), and standard SISO, for both $N = 3$ and $N = 4$. The average values are computed across all scenarios described in Section~\ref{sec:sim-scenarios}. It can be noticed that the joint MIMO operation performs better than all other modes of operation. Moreover, the joint MIMO operation provides an almost {\it linear} increase in average throughput per node with respect to SISO: when $N = 3$, the throughput gain is about 277\%, whereas V-BLAST $2 \times 3$ delivers 248\%. When devices feature $N = 4$ receive antennas, HYB-C delivers a gain of 385\% over SISO, compared to 349\% achieved with V-BLAST $3 \times 4$. 

\begin{figure}[htb]
 \centering
 \includegraphics[scale=0.45]{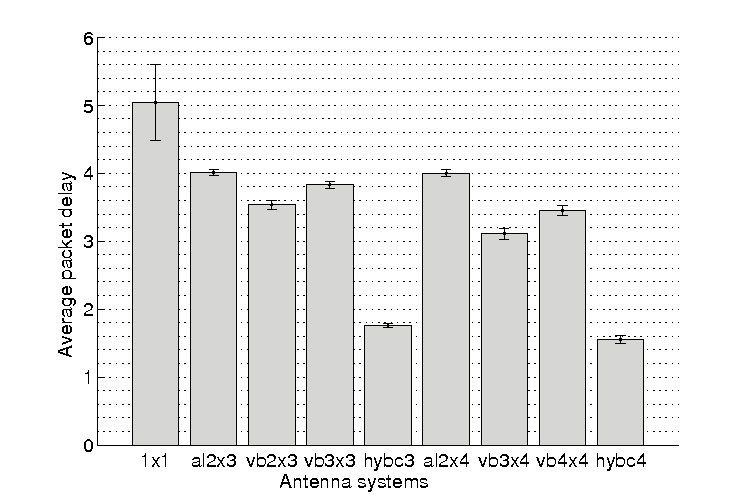}
 \caption{Average delay per frame in different modes of operation. The sufix ``al'' indicates the Alamouti scheme, ``vb'', the V-BLAST system, and ``hybc'', the HYB-C operation mode.}
 \label{fig:c_hybcdelay}
\end{figure}

Figure~\ref{fig:c_hybcdelay} depicts the {\it average delay per frame} across all topology scenarios. The results show that the joint MIMO mode of operation allows frames to be delivered under the best MIMO mode under a given $\overline{\text{SINR}}$ experienced by a given pair of nodes. Because of that, retransmissions are less frequent, and frames are delivered under considerable lower delays. In fact, the gains with respect to SISO are about 177.8\% with $N = 3$, and 233.3\% with $N = 4$. Moreover, the throughput gains with respect to V-BLAST are also significant: 94.4\% with respect to V-BLAST $2 \times 3$, and 106.7\% with respect to V-BLAST $3 \times 4$.     

\begin{figure}[htb]
 \centering
 \includegraphics[scale=0.45]{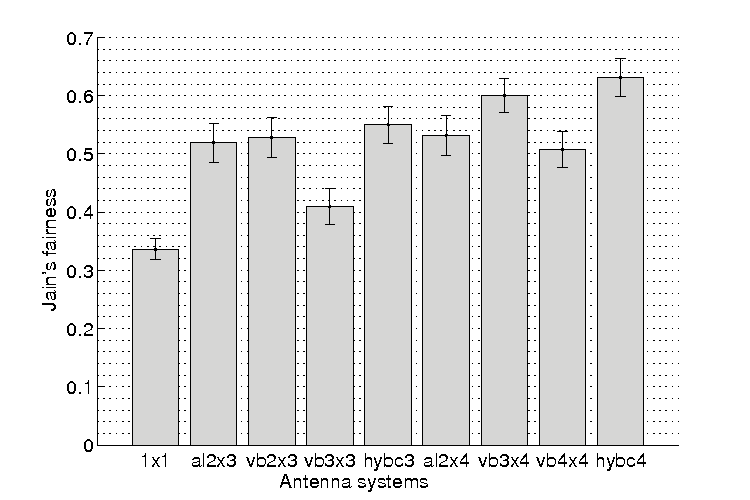}
 \caption{Throughput fairness comparision for different modes of operation. The sufix ``al'' indicates the Alamouti scheme, ``vb'', the V-BLAST system, and ``hybc'', the HYB-C operation mode.}
 \label{fig:c_hybcfairness}
\end{figure}

Finally, Figure~\ref{fig:c_hybcfairness} contains the results for {\it throughput fairness}. Compared to standard SISO, HYB-C delivers gains of about 61.8\% and 85.3\% for $N = 3$ and $N = 4$, respectively. These are significant gains, since the binary exponential backoff algorithm of the IEEE 802.11 DCF is in action under all modes of operation, and it is well known that this is the main reason for the unfairness behavior of IEEE 802.11 networks. Compared to V-BLAST, the gains of HYB-C are about 6\% ($N = 3$), and 5\% ($N = 4$), respectively.

\section{Conclusions}
\label{sec:conclusions}

This paper investigated the performance of MIMO-enabled ad hoc networks that employ the Alamouti scheme for transmit diversity, and/or the V-BLAST system for spatial multiplexing. Unlike previous works that have studied single-hop scenarios or have assumed simple abstractions to reproduce MIMO behavior, this paper considered {\it not} fully-connected MIMO ad hoc networks by taking into account the effects of multiple antennas on the CCA mechanism of CSMA-like protocols, and the impact of multiple access interference on the resulting SINR. It was shown that the average aggregate energy captured by all receive antennas must be used for CCA purposes to promote better spatial reuse. Then, based on a study about individual performance of each MIMO scheme, this paper proposes simple modifications to the IEEE 802.11 DCF MAC to allow the joint operation of both Alamouti and V-BLAST techniques, where each pair of nodes can select the best MIMO configuration for the impending data transfer. The joint operation is based on three operation modes that are selected based on the estimated SINR at the intended receiver after RTS reception and comparision with a set of threshold values. A strategy for study and selection of the theshold values was also presented for possible use in specific applications and scenarios. The performance of the joint MIMO scheme was compared with the individual MIMO schemes and the standard SISO IEEE 802.11 according to MAC-level throughput per node, delay, and fairness under saturated traffic conditions. The simulation results showed the quasi-linear increase in throughput of the proposed joint scheme, with significant gains in delay and fairness compared to the use of the individual MIMO schemes and the SISO configuration.


\section{Acknowledgments}

The authors would like to thank the financial support provided by the Coordena\c{c}\~ao de Aperfei\c{c}oamento de Pessoal de N\'ivel Superior (CAPES), Funda\c c\~ao de Amparo \`a Pesquisa do Distrito Federal (FAPDF), and the Conselho Nacional de Desenvolvimento Cient\'ifico e Tecnol\'ogico (CNPq).






\bibliographystyle{elsarticle-num}
\bibliography{relatorio}

\begin{thebibliography}{10}
\expandafter\ifx\csname url\endcsname\relax
  \def\url#1{\texttt{#1}}\fi
\expandafter\ifx\csname urlprefix\endcsname\relax\def\urlprefix{URL }\fi
\expandafter\ifx\csname href\endcsname\relax
  \def\href#1#2{#2} \def\path#1{#1}\fi

\bibitem{paul.circuitsandsystems.magazine.2008}
T.~Paul, T.~Ogunfunmi, Wireless {LAN} comes of age: Understanding the {IEEE}
  802.11n amendment, IEEE Circuits and Systems Magazine 8.

\bibitem{choudhury.icnp04}
R.~R. Choudhury, N.~H. Vaidya, Deafness: A {MAC} problem in ad hoc networks
  when using directional antennas, Proc. {ICNP} (2005) 283--292.

\bibitem{ramanathan.jsac05}
R.~Ramanathan, J.~Redi, C.~Santivanez, D.~Wiggins, S.~Polit, Ad hoc networking
  with directional antennas: A complete system solution, IEEE Journal on
  Selected Areas in Communications 23~(3) (2005) 496--506.

\bibitem{hu.jcn04}
M.~Hu, J.~Zhang, {MIMO} ad hoc networks: Medium access control, saturation
  throughput, and optimal hop distance, Journal of Communications and Networks
  (2004) 317--330.

\bibitem{ieee.levorato.casari.2007}
M.~Levorato, S.~Tomasin, P.~Casari, M.~Zorzi, Physical layer approximations for
  cross-layer performance analysis in {MIMO-BLAST} ad hoc networks, IEEE Trans.
  on Wireless Communications 6~(11).

\bibitem{rosseto.trans.wireless.comm.2009}
F.~Rossetto, M.~Zorzi, A low-delay {MAC} solution for {MIMO} ad hoc networks,
  IEEE Trans. on Wireless Communications 8~(1) (2009) 130--135.

\bibitem{babich.iet.2014}
F.~Babich, M.~Comisso, A.~Crismani, Considerations on the multiplexing and
  diversity tradeoff in {IEEE} 802.11 networks, IET Communications 8~(9) (2014)
  1551 -- 1559.

\bibitem{sundaresan.icnp.2005}
K.~Sundaresan, R.~Sivakumar, Routing in ad-hoc networks with {MIMO} links, in:
  Proc. IEEE ICNP, 2005.

\bibitem{jakllari.infocom.2006}
G.~Jakllari, S.~Krishnamurthy, M.~Faloutsos, P.~Krishnamurty, O.~Ercetin, A
  framework for distributed spatio-temporal communications in mobile ad hoc
  networks, in: Proc. IEEE INFOCOM, 2006.

\bibitem{gelal.diversity.06}
E.~Gelal, G.~Jakllari, S.~V. Krishnamurthy, Exploiting diversity gain in {MIMO}
  equipped ad hoc networks, 14th Asilomar Conference on Signals, Systems, and
  Computers (2006).

\bibitem{alamouti}
S.~Alamouti, A simple transmit diversity technique for wireless communications,
  IEEE Journal on Selected Areas in Communications 16~(8) (1998) 1451--1458.

\bibitem{ieee.foshini.ursi.1998}
P.~W. Wolniansky, G.~J. Foschini, G.~D. Golden, R.~A. Valenzuela, {V-BLAST}: an
  architecture for realizing very high data rates over the rich-scattering
  wireless channel, Proc. ISSSE (1998) 295--300.

\bibitem{ns-3}
\href{http://www.nsnam.org/}{ns-3, the network simulator}.
\newline\urlprefix\url{http://www.nsnam.org/}

\bibitem{stamoulis.transwirel03}
A.~Stamoulis, N.~Al-Dhahir, Impact of space-time block codes on 802.11 network
  throughput, IEEE Trans. on Wireless Communications 2~(5) (2003) 1029--1039.

\bibitem{xi.opnet}
W.~H. Xi, A.~Munro, M.~Barton, Link adaptation algorithm for the {IEEE} 802.11n
  {MIMO} system, in: Proc. IFIP Networking, 2008.

\bibitem{arfht.xia.hamdi}
Q.~Xia, M.~Hamdi, K.~B. Letaief, Open-loop link adaptation for next-generation
  {IEEE} 802.11n wireless networks, IEEE Trans. on Vehicular Technology 58~(7)
  (2009) 3713--3725.

\bibitem{siam.wireless.net.applications.2008}
M.~Z. Siam, M.~Krunz, Channel access scheme for {MIMO}-enabled ad hoc networks
  with adaptive diversity/multiplexing gains, Mobile Network Applications 14
  (2009) 433 -- 450.

\bibitem{siam.infocom.2006}
M.~Z. Siam, M.~Krunz, Throughput-oriented power control in {MIMO}-based ad hoc
  networks, in: Proc. IEEE ICC, 2007, pp. 3686--3691.

\bibitem{firyaguna.iccc.2012}
F.~Firyaguna, A.~C. Christ\'ofaro, E.~A. Andrade, T.~S. Bonfim, M.~M. Carvalho,
  Throughput performance of {V-BLAST}-enabled wireless ad hoc networks, in:
  Proc. IEEE ICCC, 2012, pp. 752--757.

\bibitem{andrade.camad.2013}
E.~A. Andrade, F.~Firyaguna, A.~C. Christ\'ofaro, M.~M. Carvalho, An approach
  for discrete-event simulations of alamouti scheme in ad hoc networks, in:
  Proc. IEEE CAMAD, 2013, pp. 217--221.

\bibitem{goldsmith}
A.~Goldsmith, Wireless communications, Cambridge University Press (2005).

\bibitem{mmcarvalho}
M.~M. Carvalho, J.~J. Garcia-Luna-Aceves, Analytical modeling of ad hoc
  networks that utilize space-time coding, in: In Proc. IEEE 4th Intl.
  Symposium on Modeling and Optimization in Mobile, Ad Hoc, and Wireless
  Networks (WiOpt), 2006.

\bibitem{2006.loyka.vblast.ieee.trans.comm}
S.~Loyka, F.~Gagnon, {V-BLAST} without optimal ordering: Analytical performance
  evaluation for rayleigh fading channels, IEEE Trans. on Communications 54
  (2006) 1109--1120.

\bibitem{ieeetrans.loyka.gagnon.2004}
S.~Loyka, F.~Gagnon, Performance analysis of the {V-BLAST} algorithm: an
  analytical approach, IEEE Trans. on Wireless Communications 3 (2004)
  1326--1337.

\bibitem{lacage.wns2.2006}
M.~Lacage, T.~Henderson, Yet another network simulator, in: Proc. Workshop on
  ns-2: the IP network simulator, ACM, 2006.

\end{thebibliography}

\end{document}